\documentclass[twocolumn]{aastex631}
\usepackage{acronym,hyperref}

\begin{document}

\title{First Constraining Upper Limits on Gravitational Wave Emission\\
from NS 1987A in SNR 1987A}

\author{Benjamin J. Owen}
\affiliation{
  Department of Physics and Astronomy,
  Texas Tech University,
  Lubbock, Texas 79409-1051,
  USA
}
\author{Lee Lindblom}
\affiliation{
  Center for Astrophysics and Space Sciences,
  University of California at San Diego,
  La Jolla, California 92093-0424
}
\author{Luciano Soares Pinheiro}
\affiliation{
  Department of Physics and Astronomy,
  Texas Tech University,
  Lubbock, Texas 79409-1051,
  USA
}

\begin{abstract}
We report on a search for continuous gravitational waves (GWs) from NS\,1987A,
the neutron star born in SN\,1987A.
The search covered a frequency band of 75--275\,Hz, included a wide range of
spin-down parameters for the first time, and coherently integrated 12.8\,days of
data below 125\,Hz and 8.7\,days of data above 125\,Hz from the second Advanced
LIGO observing run.
We found no astrophysical signal.
We set upper limits on GW emission as tight as an intrinsic strain of
$2\times10^{-25}$ at 90\% confidence.
The large spin-down parameter space makes this search the first astrophysically
consistent one for continuous GWs from NS\,1987A.
Our upper limits are the first consistent ones to beat an analog of the
spin-down limit based on the age of the neutron star, and hence are the first GW
observations to put new constraints on NS\,1987A.
\end{abstract}

\acrodef{GW}{gravitational wave}
\acrodef{H1}{Hanford, WA}
\acrodef{L1}{Livingston, LA}
\acrodef{O2}{Advanced LIGO's second observing run}
\acrodef{psd}{power spectral density}
\acrodef{SFT}{Short Fourier Transform}

\section{Introduction}

Due to the detection of neutrinos from the supernova \citep{Bionta1987,
Hirata1987}, it has been known since 1987 that SN\,1987A probably produced a
neutron star.
This ``NS\,1987A'' is the youngest neutron star known near our galaxy, 51.4\,kpc
away in the Large Magellanic Cloud \citep{Panagia1999}.
After many years of unsuccessful searches for a pulsar or non-pulsing neutron
star, indirect evidence has accumulated in recent years.
\citet{Cigan2019} observed infrared emission from a relatively warm, compact
blob of dust that \citet{Page2020} showed could be powered by a 30\,yr old
cooling neutron star.
\citet{Greco2021} and \citet{Greco2022} argue that the hard x-ray emission
indicates the presence of a pulsar wind nebula.

Continuous \ac{GW} emission from NS\,1987A has been suggested since
\citet{Piran1988}.
Such \acp{GW} could be produced by a nonaxisymmetric deformation of the neutron
star, by free precession, or by long-lived $r$-mode oscillations, as summarized
\textit{e.g.} by \citet{Glampedakis2018}.

The most recent search for continuous \ac{GW} emission from NS\,1987A used
stochastic background methods to analyze data from Advanced LIGO and Virgo's
first three observing runs \citep{KAGRA:2021mth}.
Like previous similar searches referenced therein, \citet{KAGRA:2021mth}
assumed a small spin-down for NS\,1987A.
Of order $10^{-9}$\,Hz/s, this spin-down is large by the standards of known
pulsars but small because it implies an enormous unphysical accretion spin-up to
counteract the spin-down implied by \ac{GW} emission at the high level required
for detection (see below).

\citet{Wette2008} scoped out broad band continuous \ac{GW} searches for
supernova remnants (such as 1987A) where there is evidence for a neutron star
but pulses are not observed, and hence a wide range of frequencies and
spin-downs (time derivatives of the frequency) must be covered.
(Continuous \ac{GW} searches involve longer coherence times than stochastic
background searches, and typically search over spin-down parameters as well as
\ac{GW} frequencies.)
The approach of \citet{Wette2008} was first used on the young neutron star in
Cas~A \citep{LIGOScientific:2010tql}, and similar methods have been used on
other likely locations of neutron stars not observed as pulsars
\citep{LIGOScientific:2013wcb, LIGOScientific:2014ymk, Zhu:2016ghk, Sun2016,
LIGOScientific:2016ave, LIGOScientific:2018esg, Dergachev:2019pgs, Ming:2019xse,
Piccinni:2019zub, Lindblom2020, Millhouse:2020jlt, Papa:2020vfz, Jones:2020htx,
LIGOScientific:2021mwx, LIGOScientific:2021inr, Ming:2021xtz}.

\citet{Wette2008} defined a key figure of merit, an indirect limit on \ac{GW}
emission similar to the spin-down limit for pulsars---\textit{i.e.,} a best case
amplitude of \ac{GW} emission.
Even when the spin-down is not known, one can assume that it has been dominated
by \acp{GW} since the birth of the star and that it has spun the star down
significantly from its birth frequency.
This results in a frequency-independent limit based on the age of the star,
extended by \citet{Owen2010},
\begin{eqnarray}
\label{h0age}
h_0^\mathrm{age} &=& 2.5\times10^{-25} \left( \frac{51.4\, \mathrm{kpc}} {D}
\right) \left( \frac{30\, \mathrm{yr}} {a} \right)^{1/2}
\nonumber\\
&& \times \left( \frac{I} {10^{45}\, \mathrm{g}\, \mathrm{cm}^2} \right)^{1/2}
\left( \frac{6} {n-1} \right)^{1/2}.
\end{eqnarray}
Here $h_0$ is a measure of \ac{GW} amplitude called the intrinsic strain
\citep{Jaranowski1998}, $D$ is the distance to the neutron star, $a$ is its age,
$I$ is its moment of inertia, and $n = f\ddot{f} / \dot{f}^2$ is its braking
index.
The braking index is about 5 or 7 if the \ac{GW} emission is due to a corotating
nonaxisymmetry (ellipticity) or an $r$-mode respectively.
The fiducial moment of inertia in Eq.~(\ref{h0age}) is on the low end of the
predicted range, and depending on the star's mass and the nuclear matter
equation of state $h_0$ could go up by about a factor 2 \citep{J0537Rmodes}, so
the range of $h_0^\mathrm{age}$ for NS\,1987A is about 2--4$\times10^{-25}.$

\citet{Sun2016} performed the most recent search for NS\,1987A that used
continuous \ac{GW} methods, using the setup of \citet{Chung2011}, and summarize
earlier searches for that star.
Continuous wave methods are generally more sensitive than stochastic background
methods but more computationally intensive.
Since the \citet{Wette2008} wide parameter space was unfeasible for a source as
young as NS\,1987A (19 years old for the data used and requiring a fourth
spin-down parameter), \citet{Chung2011} narrowed the search by introducing a
detailed spin-down model.
But this is less robust than a model that makes few assumptions like
\citet{Wette2008}, and even with a narrow parameter space \citet{Sun2016} did
not place upper limits beating the indirect limit $h_0^\mathrm{age}.$
Recent all-sky surveys for continuous \acp{GW} such as
\citet{LIGOScientific:2022pjk} and \citet{Dergachev:2022lnt} reach
$h_0^\mathrm{age}$ in the direction of NS\,1987A but cover too small a spin-down
range for it.

As of \ac{O2}, NS\,1987A was 30 years old.
For that age and a 51.4\,kpc distance \citep{Panagia1999}, $h_0^\mathrm{age}$ is
comparable to what recent \ac{GW} searches of supernova remnants such as
\citet{Lindblom2020} have achieved using only two spin-down parameters.
The results of \citet{Wette2008} and \citet{Wette2012} can be combined to
estimate that a coherent search of \ac{O2} data can use only two spin-down
parameters and surpass the sensitivity of $h_0^\mathrm{age}$ for a computing
budget of order a million core-hours on a modern cluster.

Here we describe such a search, which detected no astrophysical signals but
placed the first direct upper limits on \ac{GW} strain from NS\,1987A to beat
the indirect limit $h_0^\mathrm{age}$ over a wide and physically consistent
parameter space.

\section{Search methods}

Since our search methods were much like those of \citet{Lindblom2020} and
similar papers, we only summarize highlights and changes here and direct the
reader to \citet{Lindblom2020} and references therein for details.
Our input parameters and some derived parameters are given in Table~\ref{pars}.

\begin{table*}
\begin{tabular}{ll|lll}
\multicolumn{2}{c}{Input parameters} &
\multicolumn{3}{c}{Derived parameters}
\\
Name & Value & Name & Value (75--125\,Hz) & Value (125--275\,Hz)
\\
\hline
Right ascension & 05$^h$ 35$^m$ 28$^s$.0 & Span & 12.76\,d & 8.73\,d
\\
Declination & $-69^\circ$ 16' 11'' & Start & 2017-06-22 20:29:29 &
2017-02-11 16:29:09
\\
Age & 30\,yr & H1 SFTs & 497 & 346
\\
Distance & 51.4\,kpc & L1 SFTs & 490 & 338
\end{tabular}
\caption{
\label{pars}
Parameters used in our search.
The position was taken from the SIMBAD database.
Times are UTC.
}
\end{table*}

We used LIGO open data \citep{Vallisneri2015, LIGOScientific:2019lzm} from
\ac{O2}, the most recent data publicly available when we started our
computational runs, in the form of 1800\,s \acp{SFT}.
\ac{O2} included data from the \ac{H1} and \ac{L1} 4\,km interferometers.
Once the integration time spans for our search bands were determined by
computational cost (see below), we selected the stretch of data for each span to
maximize sensitivity, which is proportional to live time over the \ac{psd} of
strain noise \citep{Jaranowski1998}.

The integration method was the multi-detector $\mathcal{F}$-statistic
\citep{Jaranowski1998, Cutler2005}, which efficiently accounts for the
modulation of long-lived signals due to the rotation of the Earth.
Because $2\mathcal{F}$ is a quadrature of four matched filters, in stationary
Gaussian noise it is drawn from a $\chi^2$ distribution with four degrees of
freedom.
If a signal is present, the $\chi^2$ is noncentral and the amplitude
signal-to-noise ratio is roughly $\sqrt{\mathcal{F}/2}.$

We assumed that the demodulated signal frequency evolved in the solar system
barycenter frame as
\begin{equation}
\label{foft}
f(t) = f + \dot f (t-t_0) + \frac{1}{2} \ddot f (t-t_0)^2,
\end{equation}
where the reference time $t_0$ is the beginning of the observation and the
parameters $(f, \dot f, \ddot f)$ are evaluated at that time.
That is, we assumed no binary companion to NS\,1987A, no glitches during the
spans of integration, little timing noise, and insufficient frequency drift to
require a third derivative.

To choose the parameter space---\textit{i.e.,} ranges of $(f, \dot f, \ddot
f)$---we first split the search into low and high frequency bands divided at
125\,Hz, roughly twice the spin frequency of the fastest known young pulsar
\citep{Marshall1998}.
The latter should be the frequency of most efficient emission of mass
quadrupolar \acp{GW}.
For a given frequency $f,$ the ranges of $(\dot f, \ddot f)$ were chosen the
same way as in \citet{Lindblom2020}.
That is,
\begin{eqnarray}
\frac{f} {(n_{\max}-1)a} \le & -\dot f & \le \frac{f} {(n_{\min}-1)a},
\\
n_{\min} \frac{\dot f^2} {f} \le & \ddot f & \le n_{\max} \frac{\dot f^2} {f}.
\end{eqnarray}
with the braking index ranging from $n_{\min}=2$ to $n_{\max}=7.$
These ranges correspond to a wide range of observed and predicted behaviors, and
are consistent with the minimum spin-down
\begin{eqnarray}
-\dot f &=& 1.6 \times 10^{-8} \mbox{Hz/s} \left( \frac{\kappa} {2} \right)^2
\left( \frac{D} {\mbox{51.4 kpc}} \right)^2
\nonumber\\
&& \times \left( \frac{h_0} {2\times10^{-25}} \right)^2 \left( \frac{f}
{\mbox{100 Hz}} \right) \left( \frac{10^{45}\,\mbox{g cm}^2} {I} \right)
\end{eqnarray}
required for a given $h_0$---see, \textit{e.g.,} \citet{Owen2010}.
Here $\kappa$ is the ratio of \ac{GW} frequency to spin frequency.
Note that the minimum value of $-\dot f$ is greater than the maximum value
covered by all-sky surveys \citep{LIGOScientific:2022pjk}.
The minimum frequency of the low band (75\,Hz) and the maximum frequency of the
high band (275\,Hz) were chosen so that, according to the sensitivity estimate
of \citet{Wette2012}, upper limits on $h_0$ would just reach $h_0^\mathrm{age}.$
The precise value of $h_0^\mathrm{age}$ we chose for this purpose was the
intermediate one displayed in Eq.~(\ref{h0age}), for $r$-mode emission from a
low mass (and moment of inertia) star or mass quadrupole emission from an
intermediate mass star, about $2.5\times10^{-25}.$
For a computational cost of $10^6$ core-hours per band, this resulted in the
integration times and other parameters shown in Table~\ref{pars}.

The code was an improved version of that used in \citet{Lindblom2020}, based on
the \texttt{S6SNRSearch} tag of the \textsc{LALSuite} software package
\citep{LALSuite} and its implementation of the $\mathcal{F}$-statistic.
The search parameter space was split into roughly $10^5$ batch jobs per band,
each taking roughly ten hours on the Texas Tech supercomputing cluster
``Quanah.''
We also used these jobs as an \textit{ad hoc} way of clustering candidate
signals (see below).

We did not \textit{a priori} veto candidates based on time-frequency behavior or
lists of known instrumental lines.
Due to the rapid spin-down a detectable signal would have, most templates
overlap a spectral line for some time and such vetoes would render much of the
search band unusable for setting upper limits.
The rapid spin-down has the advantage, however, of diluting the effect of a
narrow line on any given template, as the template relatively rapidly moves out
of the disturbed frequency band.
This is shown by the relatively small number of candidates (see below).
We did use the interferometer consistency veto, a simple check first used in
\citet{LIGOScientific:2012pjg} that the joint $2\mathcal{F}$ is greater than the
value from either interferometer alone.

We performed consistency checks as in \citet{Lindblom2020}, plus two more
necessitated by the unusual youth of the target and hence the high value of
spin-downs searched:
We confirmed that a third frequency derivative is not needed in Eq.~(\ref{foft})
and that the standard \ac{SFT} length of 1800\,s is not too long.

Inspection of the parameter space metric \citep{Wette2008} shows that omitting
the third frequency derivative can result in a substantial mismatch between
signal and template for the parameters and integration times used here.
However \citet{Jaranowski2000} argued that correlations with lower derivatives
allow for neglect of the third derivative at much longer integration times than
we use, while still keeping a low mismatch and thus a high fraction of the ideal
$2\mathcal{F}.$
Essentially, correlations allow a large template bank to pick up the signal
efficiently at a shifted position $(f, \dot{f}, \ddot{f}).$
This argument is weak near the edges of parameter space, so we checked against a
set of software injections (with the highest third derivatives allowed by our
braking index range) and confirmed that omitting the third derivative causes no
appreciable loss in $2\mathcal{F}$ for a population of signals.
As \citet{Jaranowski2000} argue, there is no detectable effect.

One might expect the \ac{SFT} length to be a problem for frequency derivatives
high enough to send a signal through multiple \acp{SFT} within the duration of
one \ac{SFT}, \textit{i.e.}\ for $|\dot f|$ of order 1/(1800\,s)$^2$ or
$3\times10^{-7}$\,Hz/s---which is also the maximum $|\dot f|$ covered by our
search.
The injection checks of our upper limits (see below) already test this to some
extent, but we performed additional injection studies dedicated to this issue.
We found no significant losses for values up to $5\times10^{-7}$\,Hz/s, well
beyond what we searched.

\section{Search results}

We examined the search results for candidate signals surpassing 95\% confidence
in Gaussian noise, corresponding to $2\mathcal{F}$ thresholds of 77.1 and 77.5
for the low and high frequency bands respectively.
(These values were determined using an effective number of independent templates
found by a Kolmogorov-Smirnov distance minimizer between the observed
distributions of loudest events per search job and the Gaussian noise
prediction.)
The high band produced no candidates.
The low band produced 29 search jobs with candidates mostly clustered around
83.32\,Hz and 100.0\,Hz, with single jobs at 107.1\,Hz and 108.5\,Hz.
The highest $2\mathcal{F}$ was 95.
We examined the search jobs as in \citet{Lindblom2020} and found that all
$2\mathcal{F}$ histograms and frequency plots showed contamination by broad
noise lines.
This was sufficient to rule out the candidates as astrophysical signals, but we
followed up by checking against lists of known instrumental artifacts
\citep{Covas2018}.
The 83.32\,Hz and 100.0\,Hz clusters are due to known lines in \ac{L1} and
\ac{H1} respectively, and the narrower cluster at 108.5\,Hz is due to a known
line in \ac{H1}.
The narrower cluster at 107.1\,Hz has $2\mathcal{F}$ improbably dominated by
\ac{H1}.
The large clusters produced up to 0.1\% candidates triggering the interferometer
consistency veto, several orders of magnitude above typical search jobs.
Therefore we do not claim any astrophysical signals in our search.

\begin{figure*}
  \begin{center} 
    \includegraphics[height=0.35\textwidth]{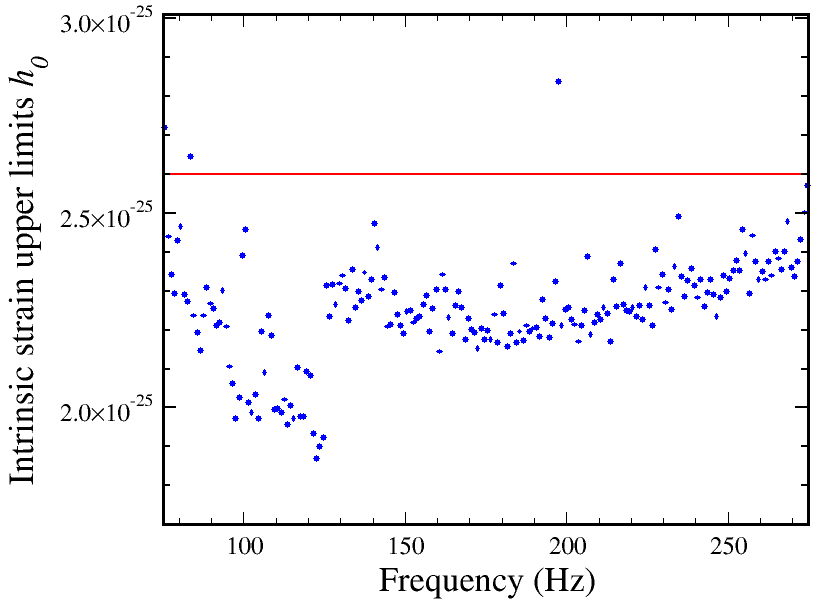}
    \hspace{0.25cm}
    \includegraphics[height=0.35\textwidth]{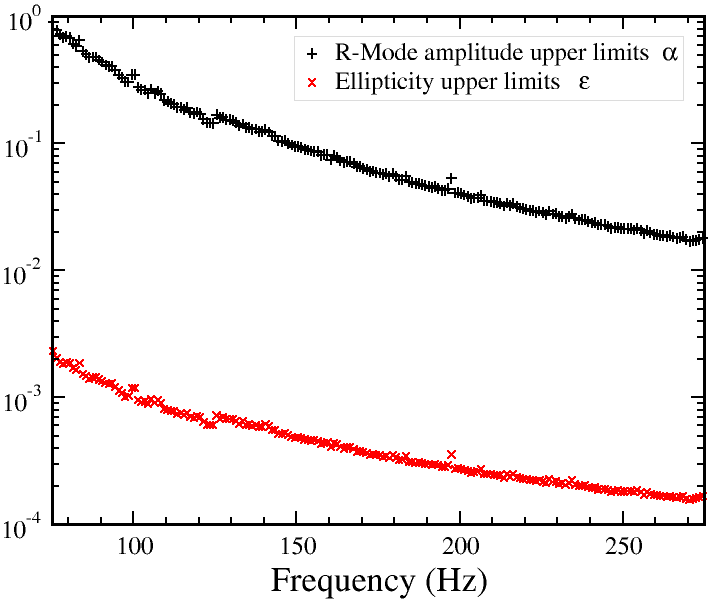}
    \hspace{0.25cm}
  \end{center}
\caption{\label{f:Fig1}
The left panel displays points representing direct observational 90\% confidence
upper limits on the intrinsic strain $h_0$ from NS\,1987A as a function of
frequency in 1\,Hz bands for our search.
The (red) horizontal line in the left panel indicates the indirect limit
$h_0^\mathrm{age}$ from energy conservation.
The right panel shows derived upper limits on the fiducial dimensionless
neutron-star ellipticity, $\epsilon$, and $r$-mode amplitude, $\alpha$, based on
these $h_0$ upper limits.
}
\end{figure*}

In the absence of signals, we set new astrophysically meaningful upper limits
on intrinsic strain $h_0$ as a function of \ac{GW} frequency in 1\,Hz bands.
The method was the same as in \citet{Lindblom2020}, with the same false
dismissal rate of 10\% (90\% confidence) integrated over a population of sources
with randomly oriented spin axes:
A semi-analytic estimate was confirmed with 1000 software-injected signals per
upper limit band.

The left panel in Fig.~\ref{f:Fig1} displays our upper limits on $h_0$ as a
function of frequency, minus two bands starting at 222\,Hz and 264\,Hz which the
injections indicated had slightly more than 10\% false dismissal rate.
The discontinuity at 125\,Hz is due to the difference in integration times used
above and below that frequency.
The (red) horizontal line in the left panel represents the fiducial value
$2.5\times10^{-25}$ of the indirect limit $h_0^\mathrm{age}$ from energy
conservation.
Our search places limits on \ac{GW} emission from NS\,1987A that are
significantly better than this astrophysical upper limit over the band searched,
and better than the strictest $h_0^\mathrm{age}$ of $2.0\times10^{-25}$ over
part of the band.

The efficiency of our search can be expressed in terms of two quantities derived
from $h_0.$
One common measure \citep{Wette2008} is the factor $\Theta$ in
\begin{equation}
h_0 = \Theta \sqrt{S_h / T},
\end{equation}
where $S_h$ is the harmonic mean \ac{psd} of all \acp{SFT} and $T$ is the total
live time of data used.
Our $\Theta$ is about 36 in the low frequency band and 40 in the high frequency
band, slightly worse (higher) than the youngest supernova remnant searches so
far---see \citet{Lindblom2020} for example.
The sensitivity depth, defined by \citep{Behnke2015} as
\begin{equation}
\mathcal{D} = \sqrt{S_h / h_0},
\end{equation}
is about 36\,Hz$^{-1/2}$ and 29\,Hz$^{-1/2}$ for the low and high bands
respectively.
This is also slightly worse (lower) than the youngest supernova remnant searches
so far, as expected due to the extreme youth and wide parameter space of
NS\,1987A.

Upper limits on $h_0$ can be converted to upper limits on fiducial neutron star
ellipticity $\epsilon$ using \citep[e.g.]{Wette2008}
\begin{equation}
\epsilon \simeq 9.5\times10^{-5} \left( \frac{h_0} {1.2\times10^{-24}} \right)
\left( \frac{D} {\mbox{1 kpc}} \right) \left( \frac{\mbox{100 Hz}} {f}
\right)^2,
\end{equation}
and to upper limits on a particular measure of $r$-mode amplitude,
$\alpha$, \citep{Lindblom1998} using \citet{Owen2010},
\begin{equation}
\alpha \simeq 0.028 \left( \frac{h_0} {10^{-24}} \right) \left( \frac{\mbox{100
Hz}} {f} \right)^3 \left( \frac{D} {\mbox{1 kpc}} \right).
\end{equation}
The numerical values are uncertain by a factor of roughly two or three due to
uncertainties in the unknown neutron star mass and equation of state.
Upper limits on $\epsilon$ and $\alpha$ for NS\,1987A from this search are shown
in the right panel of Fig~\ref{f:Fig1}.
Indirect limits on $\epsilon$ and $\alpha$ derived from $h_0^\mathrm{age}$ are
not shown since, on this logarithmic scale, they are close to the direct
observational limits.

Our upper limits on $h_0$ beat the indirect limit $h_0^\mathrm{age}.$
The latter limit is astrophysically interesting despite the youth of the source.
Equation~(\ref{h0age}) is derived under the assumption that the star has spun
down significantly since birth.
Without that assumption, but with constant braking index $n,$ the limit
$h_0^\mathrm{age}$ is multiplied by \citep{Sun2016}
\begin{equation}
\left[ 1 - \left( f_b / f \right)^{1-n} \right]^{1/2},
\end{equation}
where $f_b$ is the \ac{GW} frequency at birth.
If $f \ll f_b,$ this factor is 1 and we recover Eq.~(\ref{h0age}).
If $f_b \approx f$ as assumed by \citet{Sun2016}, essentially imposing a smaller
ellipticity, this factor can be much less than 1, but it does not need to be.
It is straightforward to insert our direct limits on $\epsilon$ and $\alpha$ and
integrate $\dot f$ to find that our assumptions are consistent---\textit{i.e.,}
that these values of $\epsilon$ and $\alpha$ result in significant spin-down
over 30 years and hence Eq.~(\ref{h0age}) is valid despite the youth of
NS\,1987A.
Thus our search had a chance of detecting a signal, and the lack of detection
represents the first \ac{GW} observational constraints on NS\,1987A.

\section{Conclusions}

We have performed the first search for \acp{GW} from NS\,1987A that covered a
physically consistent range of spin-downs and achieved a sensitivity better than
the indirect limit from energy conservation.
We also showed that this limit is applicable to NS\,1987A despite the youth of
the source.
While we detected no astrophysical signal, we set direct observational upper
limits that beat the indirect limit and thus for the first time constrain the
\ac{GW} emission of NS\,1987A if it is emitting within the frequency band
searched.
Our constraints on $r$-mode amplitude are not competitive with the standard
theoretical prediction \citep{Bondarescu2009}, but our constraints on
ellipticity are within the predicted range of elastic deformations of quark
stars \citep{Owen2005}.
If NS\,1987A is made of baryonic matter and the protons in its core are not yet
superconducting, our constraints imply upper limits on the internal magnetic
field of about $10^{15}$\,G for the twisted torus configuration likely formed
with the neutron star \citep{Ciolfi2013}.
If the protons are now superconducting, the field is likely mainly poloidal and
our constraints limit the field to less than about $10^{16}$\,G
\citep{Lander2014}.

Our search achieved this with a simple coherent integration of \ac{O2} data.
Better methods and data are available, and we expect this will motivate further
searches and improvements of search methods for rapidly evolving continuous wave
signals.

\begin{acknowledgments}

This research has made use of data, software and/or web tools obtained from the
Gravitational Wave Open Science Center (https://www.gw-openscience.org), a
service of LIGO Laboratory, the LIGO Scientific Collaboration and the Virgo
Collaboration.
LIGO is funded by the U.S. National Science Foundation. Virgo is funded by the
French Centre National de Recherche Scientifique (CNRS), the Italian
Istituto Nazionale della Fisica Nucleare (INFN) and the Dutch
Nikhef, with contributions by Polish and Hungarian institutes. This
research was supported in part by NSF grants PHY-2012857 to the University
of California at San Diego and PHY-1912625 to Texas Tech
University. The authors acknowledge computational resources provided
by the High Performance Computing Center (HPCC) of Texas Tech
University at Lubbock (http://www.depts.ttu.edu/hpcc/).
We are grateful to Katharine Long and Joseph Romano for comments on the
manuscript.

\end{acknowledgments}

\bibliography{sn1987a}

\end{document}